        \documentclass[10pt]{article}
        \usepackage[dvips]{graphicx,epsfig}
        \usepackage[usenames,dvipsnames]{color}
        \usepackage{natbib}
        \usepackage{times,mathptm,mathptmx}
        \usepackage{bm}
        \usepackage{geometry}
        \usepackage{pdflscape}
\usepackage{graphicx}
        
\begin{document}

\vspace*{1.0in}

\begin{center}

	{\large		Top-shaped Asteroids as Lens-shaped Bodies}

\vspace*{1.0in}
                        Anthony R. Dobrovolskis$^{\rm a,b,1}$, Jack J. Lissauer$^{\rm b,*}$ and Jose L. Alvarellos$^{\rm c}$
\bigskip

                        Email: Jack.Lissauer@nasa.gov

\vspace*{1.0in}
                                2023 May
\end{center}

\vspace*{1.0in}

24 pages, including

2 tables and 

12 figures (2 in color)
\bigskip

$^{\rm 1}$ Deceased 
\bigskip

$^{\rm *}$ Corresponding Author
\bigskip

$^{\rm a}$ SETI Institute, Mountain View, CA

$^{\rm b}$ Space Science \& Astrobiology Division, MS 245-3, NASA Ames Research Center, Moffett Field, CA 94035-1000

$^{\rm c}$ Flight Dynamics, Emergent Space Technologies, MS N202-3, NASA Ames Research Center, Moffett Field, CA 94035
\newpage

\vspace*{1.0in}

Proposed running head:

\begin{center}
			Top-shaped Asteroids
\end{center}

\vspace*{1.0in}

Correspondence:

\begin{center}
                        Jack J. Lissauer

                        245-3 NASA Ames Research Center

                        Moffett Field CA 94035-1000

                        Jack.Lissauer@nasa.gov

                        'Phone: (650) 229-3990

                        FAX: (650) 604-6779

\end{center}

\vspace*{0.3in}


Asteroids, rotation

Asteroids, surfaces

Celestial Mechanics



Rotational Dynamics


\newpage

\vspace*{0.5in}

HIGHLIGHTS \\







- Top-shaped asteroids (TSAs) are modeled as symmetric lens-shaped bodies 

- Their total energy E (self-gravitational plus rotational) is found analytically 

- The lens shape of lowest E is found for a given mass, density, and angular momentum 

- Known TSAs do not conform to the lens shape of lowest E

- Other processes must control the shapes of TSAs

\vspace*{0.5in}

\begin{center}
                                ABSTRACT
\end{center}

Several asteroids are known to be shaped like toy tops.  This paper 
models Top-Shaped Asteroids (TSAs) as Homogeneous Symmetric Lenses (HSLs), 
and derives their rotational, self-gravitational, and total energies 
as functions of their mass, density, and angular momentum.  
Then we raise, test, and ultimately reject the hypothesis 
that TSAs take the shape of lowest total energy, 
subject to the constraint that they keep the same mass, 
density, and angular momentum, while remaining HSLs.  
Other processes must control the shapes of TSAs.  
For completeness,  we also describe a Core-Mantle Model for TSAs, 
as well as an Inverted Core-Mantle Model, 
and derive their self-gravitational energies, 
along with their rotational energies.  
The gravitational potential at the center of an HSL then is derived.

\newpage

\section{Introduction}

\begin{figure}
\includegraphics[width=5.25in]{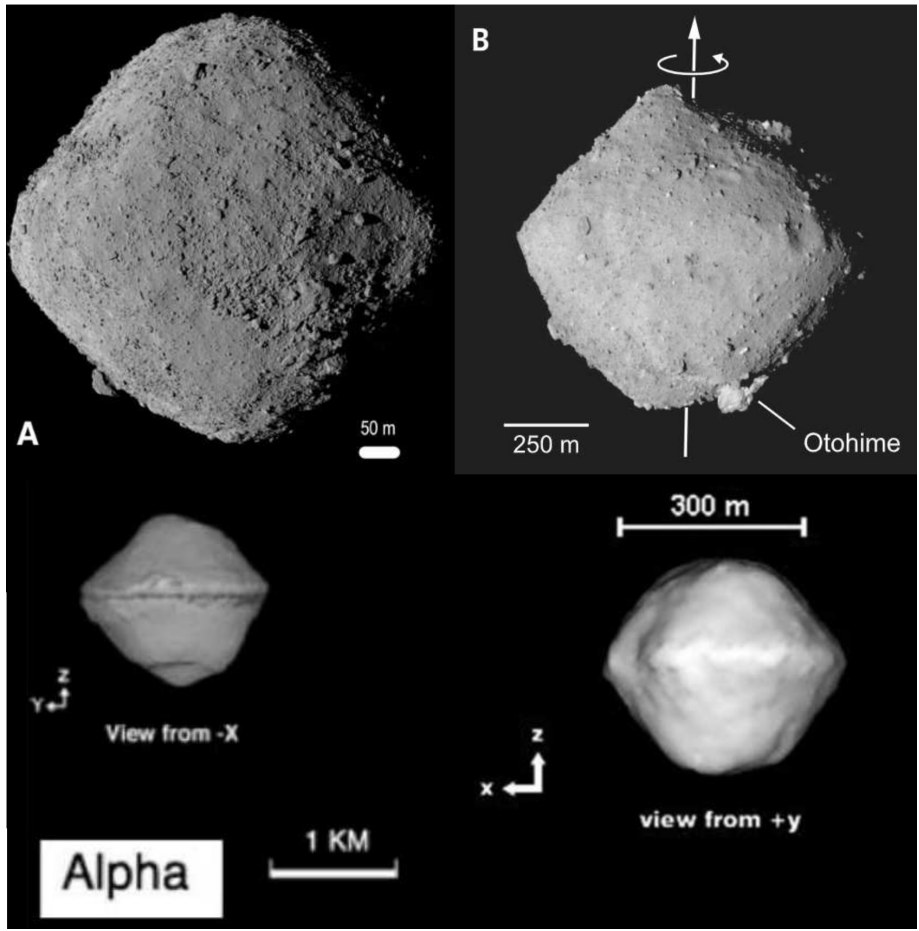}
\caption{ Four Top-Shaped Asteroids (not to scale).  
Upper left (panel A):  Visible-light image of (101955) Bennu, 
adapted from Fig. 1 of Daly {\it et al.} (2020).  
Upper right (panel B):  Visible-light image of (162173) Ryugu, 
adapted from Fig. 1 of Watanabe {\it et al.} (2019).  
Bottom left:  Radar shape model of (66391) 1999 KW4 Alpha, 
adapted from Fig. 3 of Ostro {\it et al} (2006).  
Bottom right:  Radar shape model of (341843) 2008 EV5, 
adapted from Fig. 5 of Busch, Ostro, {\it et al} (2011).  }
\end{figure}





Several asteroids are known to be shaped like toy tops (see Fig. 1).  
This paper models Top-Shaped Asteroids (TSAs) 
as Homogeneous Symmetric Lenses (HSLs), 
consisting of two identical homogeneous spherical caps of height $h$ 
sharing a common circular base of radius $R_E$ (see Fig. 2).  
The dimensionless aspect ratio is defined by $\eta \equiv h/R_E$.  
Symbols and some simple formulae are listed in Table 1.  

\begin{figure}
\includegraphics[width=5.25in]{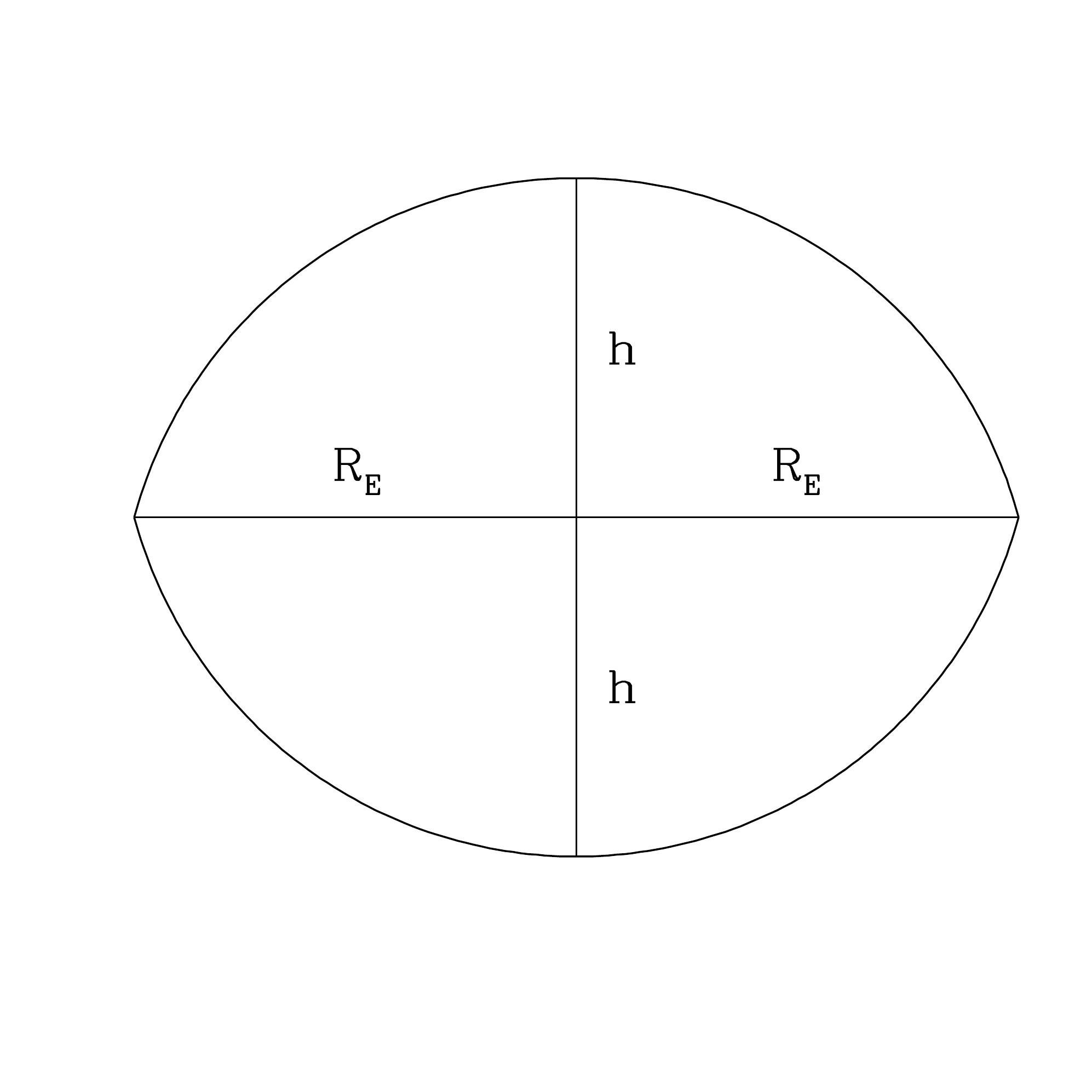}
\caption{ Polar cross-section of a Homogeneous Symmetric Lens 
with equatorial radius $R_E$ and polar radius $h$.  This sketch 
is drawn for an aspect ratio $\eta \equiv h/R_E$ = 0.767 327, 
so that the interior angle at the equatorial cusp is 150$^\circ$.  }
\end{figure}

Each ``hemisphere'' of a TSA actually resembles a paraboloid, a hyperboloid, 
or even the frustum of a cone more than a spherical cap.  However, although 
the external gravitational fields of all these shapes are still unknown, 
the self-gravitational energies of spherical lenses are known 
(Kondrat'ev, 1993; Kondrat'ev and Antonov, 1993; Seidov, 2000a,b), 
while those of paraboloids, hyperboloids, and frusta are not.  

Once the self-gravitational energy $E_G$ is known, 
it can be added to the rotational energy $E_R$ 
to find the total energy $E = E_R +E_G$ of an HSL 
as a function of its mass, dimensions, and rotation rate.  
This paper raises the hypothesis that TSAs take the shape of lowest total energy $E$, 
subject to the constraint the they keep the same mass, density, and angular momentum, 
while remaining HSLs.  

\begin{center}
                                        Table 1.  Symbols.
\vspace*{0.1in}

\begin{tabular}{|c|c|}
\hline
		$\delta$	&			density of core minus density of mantle			\\
		$\eta$		&			dimensionless aspect ratio $h/R_E$			\\
		$\lambda$	&				latitude					\\
		$\rho$		&			mass density of HSL					\\
		$\phi$		&				longitude					\\
	$\boldsymbol{\omega}$	&			spin angular velocity					\\
\hline
\hline
		$A,B$		&			equatorial moments of inertia				\\
		$a,b$		&			equatorial equivalent radii				\\
		$C$		&			polar moment of inertia					\\
		$c$		&			polar equivalent radius					\\
\hline
		$E$		&			total energy = $E_G +E_R$				\\
		$E_G$		&			self-gravitational energy				\\
		$E_R$		&			rotational energy = $J^2/(2C)$				\\
\hline
		$G$		&			Newton's constant of universal gravitation		\\
		$h$		&		height of spherical cap = $R_C -\sqrt{R_C^2 -R_E^2}$		\\
		$\bf J$		&			spin angular momentum = $C\boldsymbol{\omega}$		\\
		$M$		&			mass of TSA = $\rho V$					\\
		$M_S$		&			mass of central sphere = $4\pi\delta R_S^3/3$		\\
\hline
		$\bf r$		&			location of field point from center			\\
		$R_C$		&			radius of curvature = $(R_E^2 +h^2)/(2h)$		\\
		$R_E$		&			equatorial radius = $\sqrt{2hR_C -h^2}$			\\
		$R_S$		&			radius of homogeneous spherical core			\\
		$R_V$		&	radius of a sphere with volume $V$:  $R_V = (\frac{3V}{4\pi})^{1/3}$	\\
\hline
		$S$		&			surface area = $4\pi hR_C$				\\
		$U$		&			gravitational potential					\\
		$U_0$		&			gravitational potential at center			\\
		$V$		&			volume = $\pi h[R_E^2 +h^2/3]$				\\
		$W$		&			dimensionless parameter describing rotation		\\
\hline

\end{tabular}

\end{center}

Our model for top-shaped asteroids is presented in Section 2. The next three sections derive properties of this model. Our model is applied to observed top-shaped asteroids in Section 6. The primary conclusion from our modeling effort regarding the cause of asteriods being top-shaped is given in Section 7. Appendix A derives the greatest moment of inertia of an HSL. Appendix B describes a Core-Mantle Model for TSAs, 
as well as an Inverted Core-Mantle Model, 
and derives their self-gravitational energies, 
along with their rotational energies and the gravitational potential at the center of an HSL.

\section{Model}

Each spherical cap of an HSL is of uniform mass density $\rho$
and radius of curvature $R_C = (R_E^2 +h^2)/(2h) \ge R_E$.  
The surface area $S$ of an HSL is $4\pi hR_C = 2\pi[R_E^2 +h^2] = 2\pi R_E^2 [1 +\eta^2]$,
while its volume $V$ is $2\pi h^2[R_C -h/3] = \pi h [R_E^2 +h^2/3] = \pi R_E^3[\eta +\eta^3/3]$.
Note how these reduce to the formulae $S = 4\pi R_C^2$
and $V = 4\pi R_C^3/3$ for a sphere, when $h = R_E = R_C$ and $\eta$ = 1.
Furthermore, $S = 8\pi R_C^2$ and $V = 8\pi R_C^3/3$
for two identical tangent spheres, when $h = 2R_C$ and $R_E$ vanishes.

We define the volume-equivalent radius $R_V$ of an HSL 
as the radius of a sphere with the same volume $V = 4\pi R_V^3/3$.  Then 
\begin{equation}
R_V = (h^2[3R_C -h]/2)^{1/3} = (h[3R_E^2 +h^2]/4)^{1/3} = R_E[(3\eta +\eta^3)/4]^{1/3} .
\end{equation}
Note that $h \le R_V \le R_E$.  With a little trigonometry, 
the interior angle at the equatorial cusp works out to be 
\begin{equation}
                180^\circ -2{\rm Arcsin}(\frac{1-\eta^2}{1+\eta^2})
                        = 2{\rm Arctan}(\frac{2\eta}{1-\eta^2}).
\end{equation}

For a sphere, when $\eta$ = 1, note how $R_V$ reduces to $R_E$, while 
the interior angle becomes $180^\circ$, and the equatorial cusp vanishes.  
Henceforth, we restrict $\eta$ to be less than or equal to unity; 
then $R_V \le R_E \le R_C$, and $\eta \le 180^\circ$.  
Figure 3 graphs the interior angle at the equatorial cusp from Formula (2) above, 
as a function of the dimensionless aspect ratio $\eta \equiv h/R_E$.  

\begin{figure}
\includegraphics[width=5.25in]{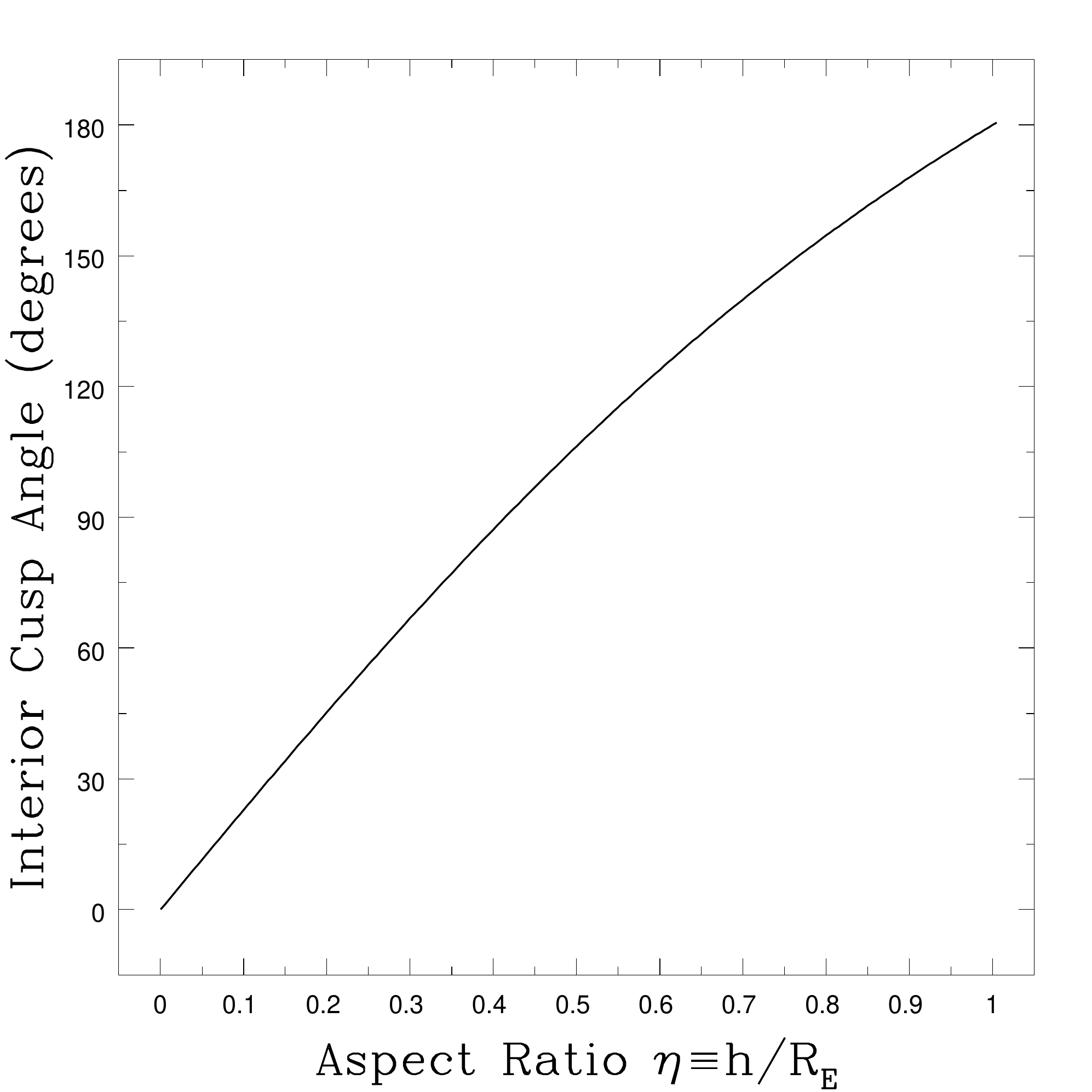}
\caption{ Interior angle at the cusp of a Homogeneous Symmetric Lens, 
as a function of its dimensionless aspect ratio $\eta \equiv h/R_E$. }
\end{figure}

\newpage

\section{Rotation}

Essentially all celestial bodies are spinning.  
Because internal stresses and strains dissipate rotational energy, 
imperfectly elastic objects such as comets, asteroids, and satellites 
eventually spin about their central axes of greatest moment of inertia $C$, 
unless forced into a state of tumbling (non-principal axis rotation) 
by non-gravitational forces, or into chaotic rotation by tides.  

For a Homogeneous Symmetric Lens, this spin axis is its axis of symmetry, 
and its center of mass lies at its geometric center.  
Its greatest moment of inertia works out to be 
\begin{equation}
		C = \frac{\pi\rho h}{30}[10R_E^4 +5R_E^2 h^2 +h^4] , 
\end{equation}
while its equatorial moments of inertia are 
\begin{equation}
		A = B = \frac{\pi\rho h}{60}[10R_E^4 +15R_E^2 h^2 +7h^4]
\end{equation}
		(Levinson, 2010; see also Appendix A).  

It is easy to verify that $A \le B \le C$, but $A +B \ge C$, as required.  
Furthermore, as $\eta$ approaches unity, $h$ approaches $R_E$, 
and $A$, $B$, and $C$ all become $8\pi\rho R_E^5/15 = 2MR_E^2/5$ for a sphere.  

To illustrate, Fig. 4 graphs $A$, $C$, and $A+B = 2A$ of an HSL 
from Formulae (3) and (4) above, all normalized by $MR_V^2$, 
as functions of the dimensionless aspect ratio $\eta \equiv h/R_E$.  
Note how $C$ always lies between $A$ and $A+B = 2A$, as required.  
For small $\eta$, $C \approx 2A$, but all three curves approach infinity.  
At $\eta$ = 1, $C = A = 2MR_V^2/5$, as expected for a sphere.  

\begin{figure}
\includegraphics[width=5.25in]{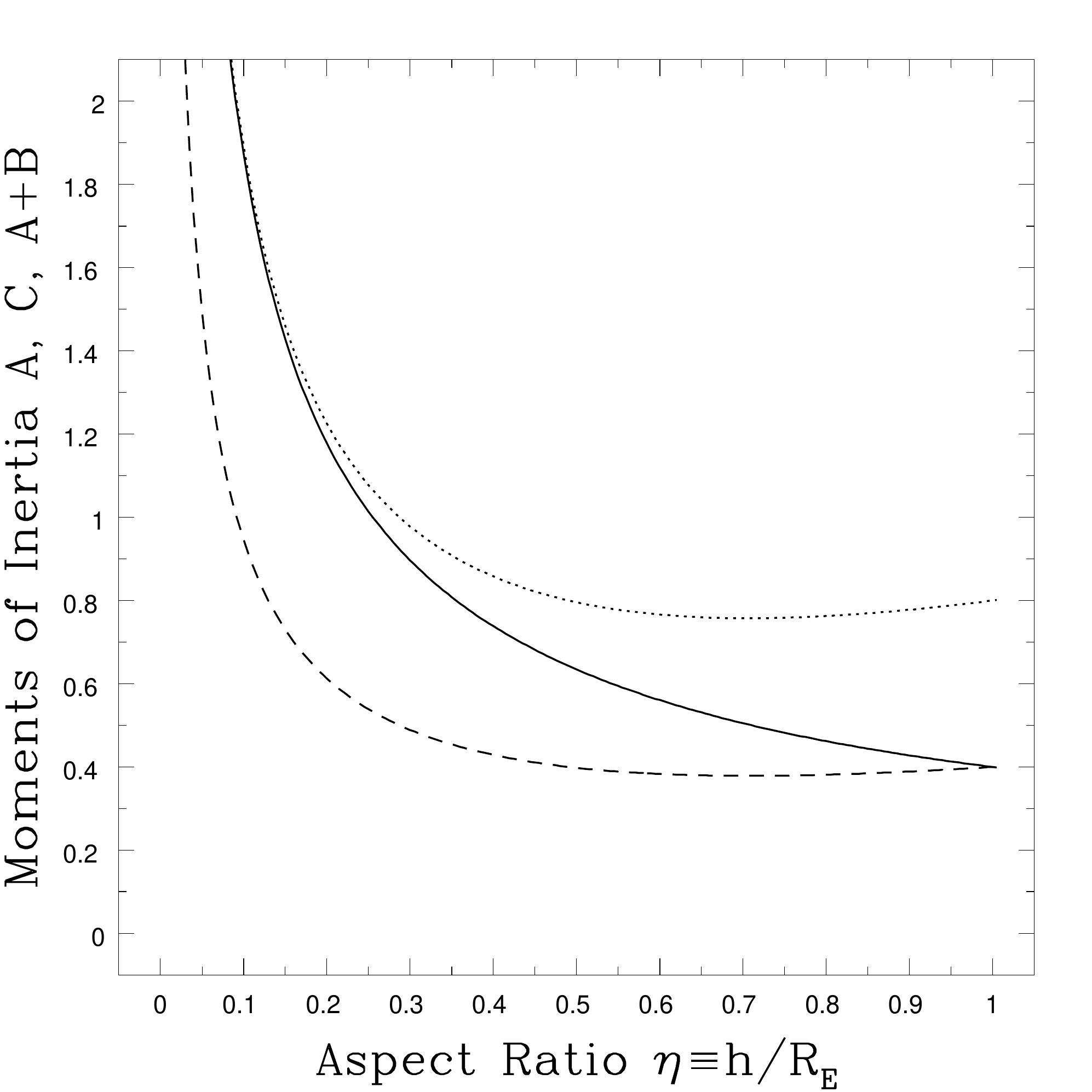}
\caption{ Moments of inertia $A$, $C$, and $A+B = 2A$ of a Homogeneous Symmetric Lens,
all normalized by $MR_V^2$, as a function of its dimensionless aspect ratio $\eta \equiv h/R_E$.  
Solid curve:  Polar moment of inertia $C$.  
Dashed curve:  Equatorial moment of inertia $A = B$.  
Dotted curve:  $2A$.  }
\end{figure}

Although the normalized value of $C$ decreases monotonically with $\eta$, 
the normalized values of $A$ and $2A$ pass through an interesting shallow 
minimum at $\eta \approx$ 0.705, where $A \approx$ 0.378 74 $MR_V^2$, 
$C \approx$ 0.503 $MR_V^2$, $2A \approx$ 0.757 48 $MR_V^2$, 
and $E_G \approx$ -0.594 $GM^2/R_V$.  


The inertia tensor of an HSL is the same as that of an ``inertially equivalent ellipsoid'' 
with the same mass, and principal radii 
\begin{equation}
		a = b = \sqrt{\frac{5C}{2M}} \; \; {\rm and} \; \; c =\sqrt{\frac{5(2A-C)}{2M}}  
\end{equation}
(Dobrovolskis, 1996).  Figure 5 graphs $a$ and $c$ from Formula (5) above, 
normalized by $R_V$, as a function of the dimensionless aspect ratio $\eta \equiv h/R_E$.  
Interestingly, $a$ and $c$ turn out to lie between $\sim$ 91 \% and 100 \% 
of $R_E$ and $h$, respectively; but their ratio $c/a$ (not shown) is almost, 
but not quite, equal to $\eta$ (specifically, $\eta \le c/a <\sim$ 1.099 $\eta$).  

\begin{figure}
\includegraphics[width=5.25in]{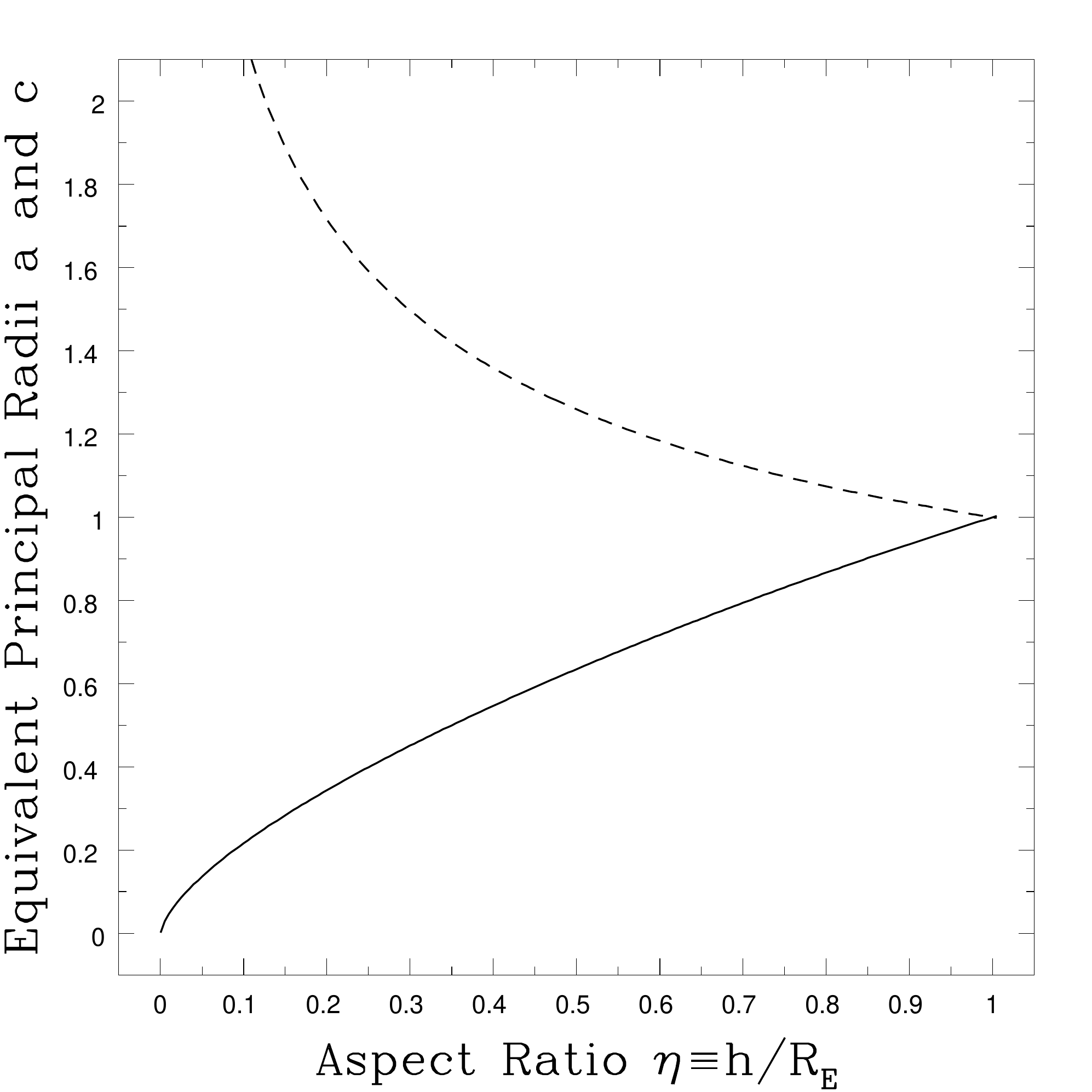}
\caption{ Inertially equivalent radii $a$ (dashed) and $c$ (solid) 
of a Homogeneous Symmetric Lens, both normalized by $R_V$, 
as functions of its dimensionless aspect ratio $\eta \equiv h/R_E$. }
\end{figure}

The rotational kinetic energy of a body is 
\begin{equation}
		E_R = C\omega^2/2 = J\omega/2 = \frac{J^2}{2C} , 
\end{equation}
where $\boldsymbol{\omega}$ 
is its spin angular velocity in radians per unit time, 
and $\bf J$ = $C\boldsymbol{\omega}$ is its spin angular momentum.  
Although the rotational energy $E_R$ is subject to dissipation, 
the spin angular momentum $\bf J$ of an isolated body 
is a conserved quantity (unless the body sheds significant mass).  

Figure 6 graphs the rotational energy $E_R$ from Formula (6) above 
as well as its derivative with respect to the dimensionless aspect ratio $\eta$, $E'_R \equiv dE_R/d\eta $, 
both normalized by the invariant quantity $J^2/(MR_V^2)$, 
as functions of $\eta$.  
Note that when $\eta$ = 0, $E_R$ vanishes too, 
while $E'_R$ is formally infinite.  
However, as $\eta$ increases to unity, $E_R$ increases monotonically 
up to the limit of 1.25 $J^2/(MR_V^2)$ for a sphere, 
while $E'_R$ decreases monotonically down to $\sim$ 0.781 $J^2/(MR_V^2)$.  

\begin{figure}
\includegraphics[width=5.25in]{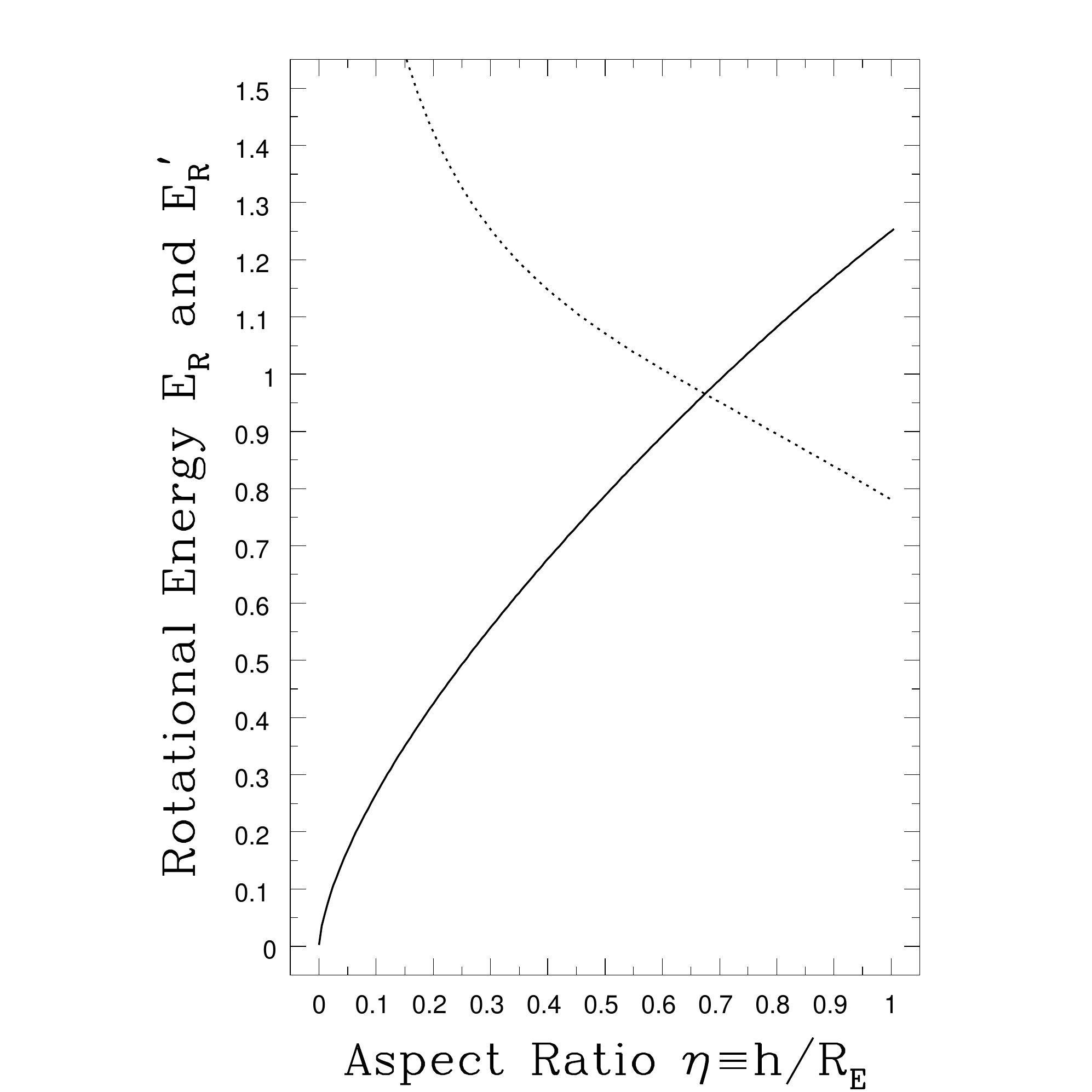}
\caption{ Rotational energy $E_R$ of a Homogeneous Symmetric Lens 
(solid curve) and its derivative $E'_R$ with respect to $\eta$ 
(dotted curve), both normalized by the invariant $J^2/(MR_V^2)$, 
as functions of the dimensionless aspect ratio $\eta \equiv h/R_E$. }
\end{figure}

\section{Gravitation}

The gravitational potential field $U({\bf r})$ of any mass distribution 
with density $\rho({\bf r})$ at any point $\bf r$ = $(x,y,z)$ can be written as 
\begin{equation}
                U({\bf r}) = -G\int\frac{\rho({\bf r^\dagger})}{|{\bf r} -{\bf r^\dagger}|}dx^\dagger dy^\dagger dz^\dagger , 
\end{equation}
where $G$ is Newton's constant of gravitation, daggered symbols denote dummy variables that are integrated over,
and the volume integral is taken over the entire mass distribution.  
Then the gravitational attraction at any point is just $-\nabla U$, 
and the gravitational potential energy of a mass element $dm$ at $\bf r$ 
is just $U({\bf r}) dm$.  

The self-gravitational energy $E_G$ of the same mass distribution 
is the potential energy it possesses by virtue of its own gravitational field.  
This is found by again integrating the mass distribution over its own potential:  
\begin{equation}
                        E_G = \frac{1}{2}\int\rho({\bf r})U({\bf r})dxdydz 
	= \frac{-G}{2}\int\int\frac{\rho({\bf r})\rho({\bf r^\dagger})}{|{\bf r} -{\bf r^\dagger}|}dxdydzdx^\dagger dy^\dagger dz^\dagger 
\end{equation}
({\it e.g.}, Kellogg, 1929; MacMillan, 1930; Ramsey, 1940).  

The leading factor of 1/2 in Formula (8) above compensates for the fact 
that the integration counts each pair of points $\bf r$ and $\bf r^\dagger$ twice.  
Note also that this paper uses the physical convention that $U$ and $E_G$ both are negative.  
Then the energy it would take to disperse the entire mass to infinity is $-E_G$.  

For example, the self-gravitational energy of a homogeneous sphere 
of radius $R_V$ is $E_G = -16\pi^2 G\rho^2 R_V^5/15$ = $-3GM^2/(5R_V)$, 
where its mass $M$ is just $\rho V = 4\pi \rho R_V^3/3$.  
Then the energy it would take to blow it up completely 
is $-E_G = +16\pi^2G\rho^2 R_V^5/15 = +3GM^2/(5R_V)$.  

Relatively few shapes have closed-form analytic (or semi-analytic) solutions for their gravitational fields; 
classically, these include the homogenous sphere (Newton, 1687), the homogeneous ellipsoid 
(Ivory, 1809; Kellogg, 1929; MacMillan, 1930; Ramsey, 1940; Danby, 1962; Kondratev, 1989, 2003, 2007), 
the uniform straight rod (e.g, Kellogg, 1929; MacMillan, 1930; Ramsey, 1940), the homogeneous polyhedron (Paul, 1974; 
Barnett, 1976; Okabe, 1979; Waldvogel, 1979; Pohánka, 1988; Werner, 1994; Broucke, 1995; Werner and Scheeres, 1995), 
and more recently, a ``duplex'' consisting of two overlapping homogeneous spheres (Dobrovolskis, 2021).  

Although the gravitational field of a HSL is not known, its self-gravitational energy is !  Kondrat’ev (1993) 
has expressed the self-energies of homogeneous bodies in terms of the spherical harmonic coefficients of their potentials.  
This method is particularly well-suited to objects with axial symmetry, such as cylinders, cones, and their frusta.  
Kondrat’ev and Antonov (1993) immediately applied this technique to derive the self-gravitational energies of homogeneous 
plano-convex lenses (segments of spheres, or spherical caps), as well as those of homogeneous biconvex lenses (see their Fig. 3).  
Later Seidov (2000a,b) generalized their result to find the self-energies of homogeneous concavo-convex lenses 
(such as a sphere with a crater shaped like a spherical bowl).

Substituting 2 Arctan$(h/R_E)$ for Arctan$(R_E/[R_C-h])$ 
from Formula (50) of Kondrat'ev and Antonov (1993) 
into the final term of their Formula (67) gives 
\begin{equation}
                E_G = -\frac{4\pi}{9} G\rho^2 \{ 10 R_C^4 R_E -\frac{16}{3} R_C^2 R_E^3 -\frac{8}{3}R_E^5 
                        +\pi h^4 [\frac{2}{5}h -2R_C -\frac{R_C^2}{R_C-h}] 
                        +4R_C^4[\frac{R_C^2}{R_C-h} -6R_C +6h]{\rm Arctan}(h/R_E) \} 
\end{equation}
for the self-gravitational energy of a HSL.  Kondrat'ev and Antonov (1993) evaluated the limiting case of two identical tangent spheres to yield the result 
$E_G = -(136\pi^2/45) G\rho^2R^5 (=-17GM^2/(10R)$, where here M is the mass of each individual sphere). 
They also evaluated (in their Formula 70) the (finite) difference between the two terms in their Formula 67 (and our Formula 9) that individually approach $\infty$ to recover the well-known result for the case of a single sphere.

Figure 7 graphs the self-gravitational energy $E_G$ from Formula (9) above 
as well as its derivative $E'_G$ with respect to $\eta$, both normalized by $GM^2/R_V$. 
Like $E_R$ and $E'_R$, when $\eta$ = 0, $E_G$ vanishes too, 
while $E'_G$ is formally infinite (though in the negative sense).  
In contrast to $E_R$ and $E'_R$, however, as $\eta$ increases to unity, 
$E_G$ {\it decreases} monotonically down to the limit of --0.6 $GM^2/R_V$ 
for a sphere, while $E'_G$ {\it increases} monotonically up to zero.  

\begin{figure}
\includegraphics[width=5.25in]{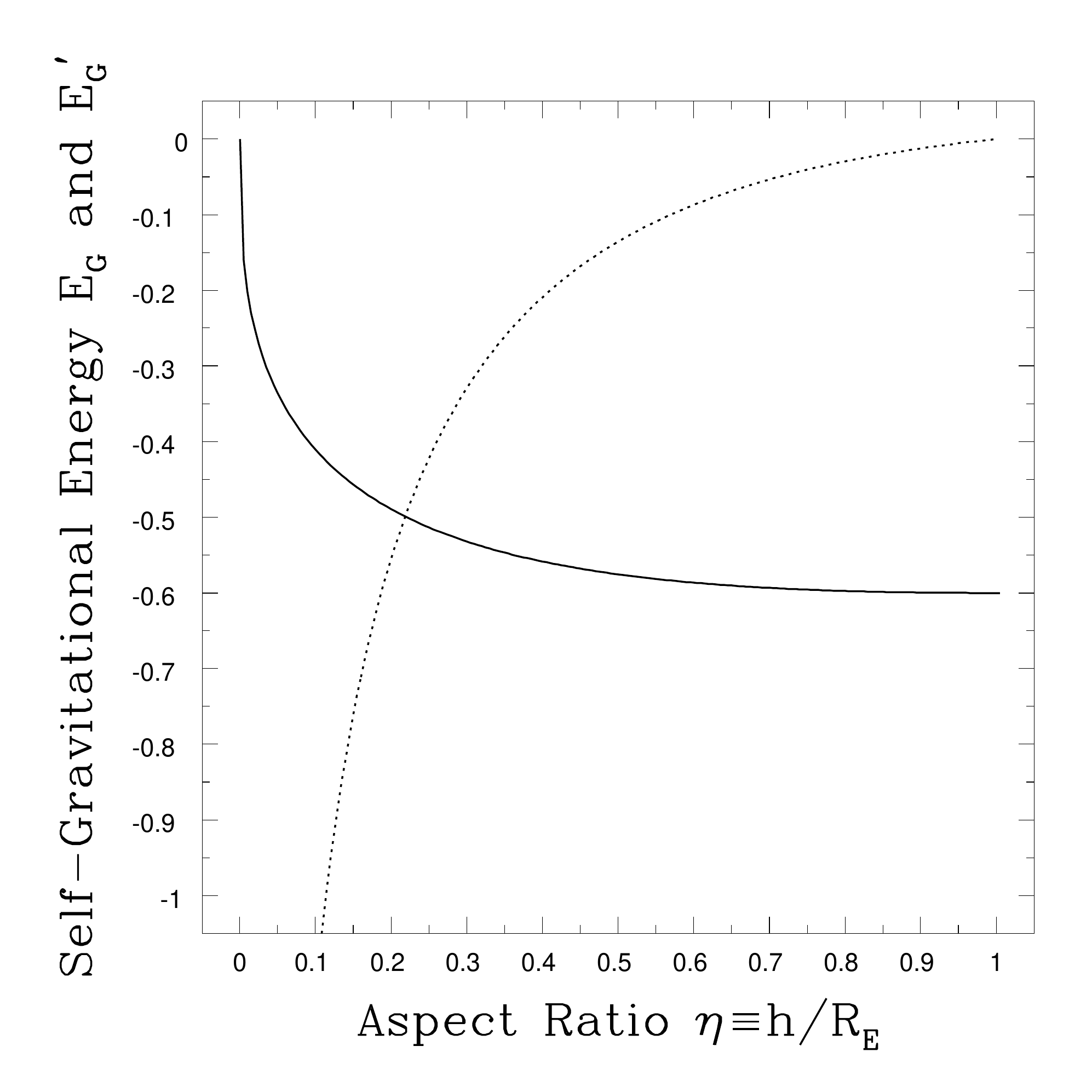}
\caption{ Self-gravitational energy $E_G$ of a Homogeneous Symmetric Lens (solid curve) 
and its derivative $E'_G$ with respect to $\eta$ (dotted curve), both normalized by $GM^2/R_V$, 
as functions of the dimensionless aspect ratio $\eta \equiv h/R_E$.  }
\end{figure}

\newpage

\section{Total Energy}

The total energy $E$ of an isolated body is just $E_R +E_G$, the sum of its 
rotational kinetic energy and its self-gravitational potential energy.  Of course, 
if the body is rotating very slowly or not at all, $E$ is dominated by $E_G$; 
but if it is spinning very rapidly, $E$ is dominated by $E_R$.  

To quantify this, write $E_R = f(\eta)J^2/(MR_V^2)$ and $E_G = g(\eta)GM^2/R_V$, where 
$f(\eta)$ and $g(\eta)$ are the dimensionless functions of $\eta$ plotted in Figs. 6 and 7, 
respectively, while $J^2/(MR_V^2)$ and $GM^2/R_V$ are their corresponding normalization constants.  

If $E_R$ and $E_G$ are equal in magnitude, then the total energy $E$ vanishes, and we find 
\begin{equation}
				W = -g(\eta)/f(\eta) , 
\end{equation}
where we define the dimensionless rotation parameter 
\begin{equation}
		W \equiv \frac{J^2}{MR_V^2} \div \frac{GM^2}{R_V} = \frac{J^2}{GM^3R_V} . 
\end{equation}

The dashed curve in Fig. 8 graphs $W_0$, 
the value of $W$ when $E$ = 0 from Formula (10), 
as a function of the dimensionless aspect ratio $\eta \equiv h/R_E$.  
Any HSL with $W$ above this curve has positive total energy, 
and is unbound, and liable to fragmentation.  
Note that $W_0$ increases without bound as $\eta$ approaches zero, 
but it attains a minimum of 0.480 = 3/5 $\div$ 5/4 for a sphere at $\eta$ = 1.  

Thus any HSL with $W <$ 0.480 has negative total energy; 
but this does not necessarily imply that it is gravitationally bound.  
For comparison, the usual stability limit for loose material not to fly off 
from the equator of a rotating sphere is given by $\omega^2 < GM/R_V^3$; 
in terms of $W$, this becomes $W <$ 0.160 = 4/25, 
one-third as large as the criterion found above.  

Now, suppose that a Homogeneous Symmetric Lens is free to change its aspect ratio $\eta$, 
but must remain an HSL, while retaining the same mass $M$ and constant density $\rho$.  
Then it will seek the shape of lowest total energy $E$ as a function of $\eta$.  
To find this shape, set to zero the derivative of the total energy with respect to $\eta$:  
\begin{equation}
		0 = E' = E'_R +E'_G \; \; \Rightarrow \; \; W = -g'(\eta)/f'(\eta) .  
\end{equation}

The solid curve in Fig. 8 graphs $W_*$, the value of $W$ when $E'$ = 0, 
as a function of the dimensionless aspect ratio $\eta \equiv h/R_E$.   
Note that like $W_0$, $W_*$ increases without bound as $\eta$ approaches zero,
but that $W_*$ always lies well below $W_0$, and vanishes for a sphere at $\eta$ = 1.  

\begin{figure}
\includegraphics[width=5.25in]{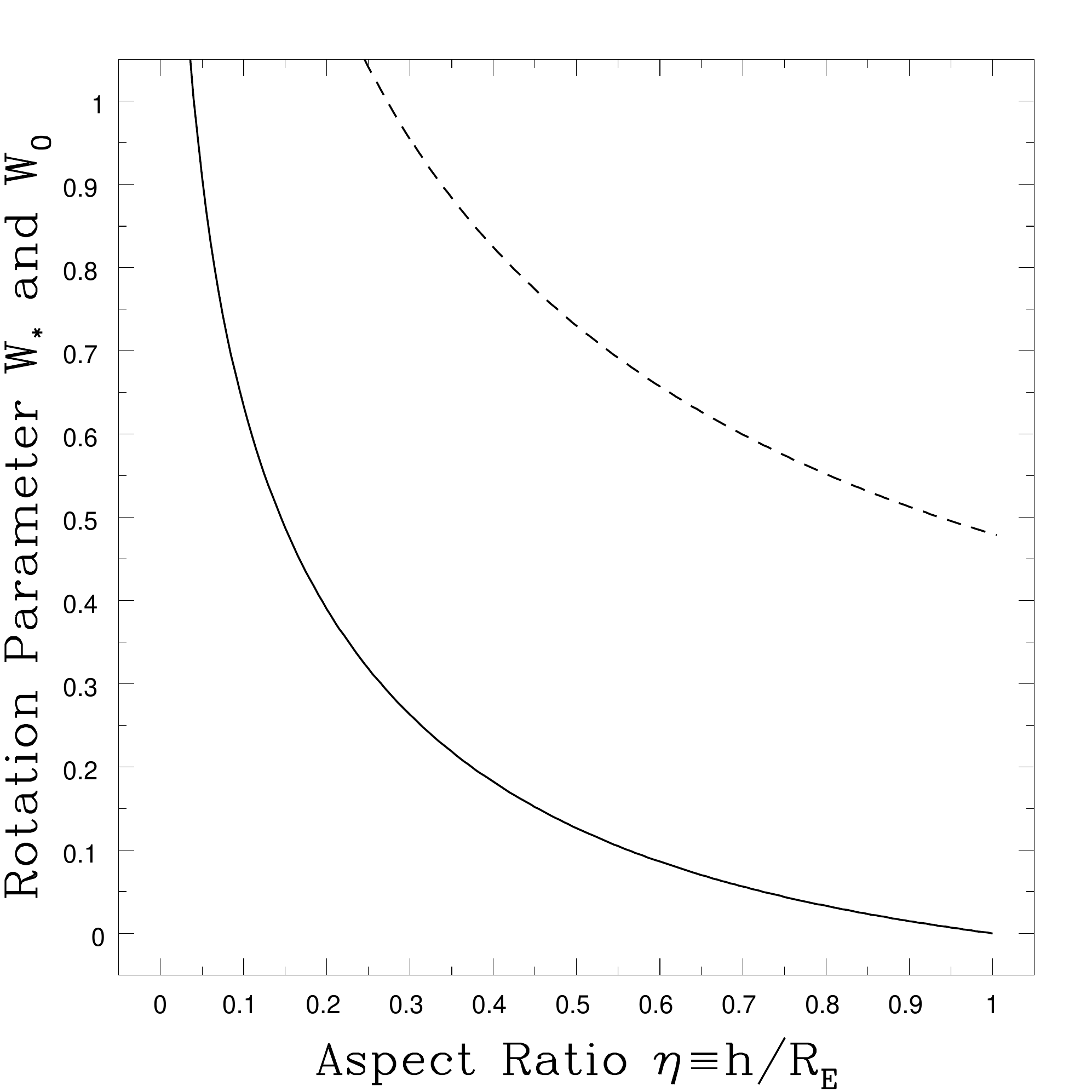}
\caption{ Relation between the dimensionless aspect ratio $\eta \equiv h/R_E$ 
of a HSL and the dimensionless ratios $W_*$ (solid curve) and $W_0$ (dashed curve) describing its rotation.  }
\end{figure}

By re-interpreting this relation as giving $\eta$ as a function of $W$,
we find the ``optimal'' shape of an HSL as a function of its rotational parameter.
For example, if $W$ = 0, the optimal shape is a sphere, not surprisingly.  
If $W$ = 0.10, in contrast, the optimal aspect ratio becomes $\eta \approx$ 0.562 .

\newpage

\section{Application to TSAs}

Table 2 lists some physical and dynamical parameters of the four TSAs depicted in Fig. 1.  
The masses of Bennu and Ryugu are obtained from the OSIRIS-REx and Hayabusa 2 orbiters, 
while the mass of 1999 KW4 Alpha is obtained from its satellite, 1999 KW4 Alpha.  
Note that the mean densities $\rho$ of Bennu and Ryugu are practically identical, 
while that of 1999 KW4 Alpha is somewhat greater.  The mass of 2008 EV5 is essentially unknown, 
but Busch, Ostro, {\it et al.} (2011) inferred a still higher density range 
of 2000 to 4000 kg/m$^3$ from the radar scattering properties of its surface.  

Note that all four aspect ratios $\eta$ are confined to a narrow range between 0.87 and 0.95; 
for comparison, the flattest planet in the Solar system is Saturn, 
an oblate spheroid with a polar radius of 0.902 times its equatorial radius.  

The moments of inertia of 2008 EV5 are directly proportional to its assumed density $\rho$, 
but its inertially equivalent radii are independent of $\rho$.  The angular momentum $J$ 
and rotational energy $E_R$ of 2008 EV5 both are proportional to $\rho$, 
while its self-gravitational energy $E_G$ is proportional to $\rho^2$, 
and its rotation parameter $W$ is inversely proportional to $\rho$.  

In all cases, the self-gravitational energy $E_G$ is several times greater in magnitude 
than the rotational energy $E_R$, so that the total energy $E$ is dominated by gravity.  
Consistent with this, $W$ is considerably less than its zero-energy value $W_0$ in each case.  

More significantly, $W$ is always much greater than its optimal value $W_*$.  
Conversely, the aspect ratio $\eta$ is always much greater than its optimal value.  
This means that these four TSAs are much more equidimensional than their relaxed states.  

\newpage
\begin{center}
			Table 2.  Parameters of four Top-Shaped Asteroids.  Those above the Reference line 
			\\ are taken from the given references; those below are derived from those above.  
\vspace*{0.1in}

\begin{tabular}{|c|c|c|c|c|c|}
\hline
					&	(101955)	&	(162173)	&	(66391)		&	(341843)	\\
		Name			&	Bennu		&	Ryugu		&    1999 KW4 Alpha	&	2008 EV5	\\
\hline
\hline
	Volume $V$ (m$^3$)		&   61.5 $\times 10^6$	&   377 $\times 10^6$	&  1 195 $\times 10^6$	&    35 $\times 10^6$	\\
	Volume-equivalent		&			&			&			&			\\
	radius $R_V$ (m)		&	246		&	448		&	658		&	200		\\
        Polar radius $h$ (m)            &       249             &       438             &       674             &       195             \\
        Equatorial radius $R_E$ (m)     &       275             &       502             &       757             &       207             \\
\hline
	Mass $M$ (kg)			&   73.3 $\times 10^9$	&   450 $\times 10^9$	&  2 353 $\times 10^9$	&	--		\\
	Mean density			&			&			&			&			\\
	$\rho = M/V$ (kg/m$^3$)		&	1 190		&	1 190		&	1 970		&	2 000 - 4 000	\\
        Rotation period (hours)         &       4.296           &       7.633           &       2.7645          &       3.725           \\
\hline
\hline
					& Lauretta {\it et al.}	& Watanabe {\it et al.}	&   Ostro {\it et al.}	&   Busch {\it et al.}	\\
		Reference		&	(2019)		&  (2019, supplement)	&	(2006)		&	(2011)		\\
\hline
\hline
        Aspect ratio $\eta \equiv h/R_E$&       0.905           &       0.873           &       0.890           &       0.942           \\
	Equatorial moments		&			&			&			&			\\
	of inertia (kg m$^2$)		& 2.40 $\times 10^{15}$	& 44.2 $\times 10^{15}$	&  600 $\times 10^{15}$	&(1.08 - 2.16) $\times 10^{15}$	\\
	Polar moment of inertia $C$	&			& 			&			&			\\
		(kg m$^2$)		& 2.62 $\times 10^{15}$	& 49.9 $\times 10^{15}$	&  666 $\times 10^{15}$	&(1.14 - 2.28) $\times 10^{15}$	\\
\hline
	Inertially equivalent 		&			&			&			&			\\
	equatorial radius $a$ (m)	&	299		&	526		&	841		&	202		\\
	Inertially equivalent		&			&			&			&			\\
	polar radius $c$ (m)		&	272		&	462		&	753		&	191		\\
\hline
        Rotation rate $\omega$ (rad/s)  &       0.000 406 3     &       0.000 228 7     &       0.000 631 34    &       0.000 468 5     \\
        Angular momentum                &                       &                       &                       &                       \\
        $J = C\omega$ (kg m$^2$/s)      &  1 065 $\times 10^9$	&  11 402 $\times 10^9$	&  42 065 $\times 10^9$	&  (53 - 107) $\times 10^9$	\\
\hline
	Rotational energy $E_R$		&			&			&			&			\\
	(kg m$^2$/s$^2$)		&   216 $\times 10^6$	&  1 300 $\times 10^6$	& 133 000 $\times 10^6$	& (130 - 250) $\times 10^6$	\\
	Self-gravitational energy $E_G$	&			&			&			&			\\
	(kg m$^2$/s$^2$)		&  --1 227 $\times 10^6$	& --22 770 $\times 10^6$	&--511 000 $\times 10^6$	& --(920 - 3700) $\times 10^6$	\\
	Total energy $E = E_R +E_G$	&			&			&			&			\\
	(kg m$^2$/s$^2$)		&  --1 011 $\times 10^6$	& --21 470 $\times 10^6$	&--378 000 $\times 10^6$	& --(800 - 3400) $\times 10^6$	\\
\hline
	Rotation parameter $W$		&	0.175		&	0.0477		&	0.309		&	0.062 -- 0.031	\\
        $W_0$ = Zero-energy value of $W$&	0.511		&	0.523		&	0.516		&	0.498		\\
	$W_*$ = Optimal value of $W$	&	0.0140		&	0.0193		&	0.0164		&	0.00822		\\
	Optimal value of $\eta$		&	0.411		&	0.734		&	0.258		&	0.678 - 0.810	\\
\hline

\end{tabular}

\end{center}

\section{Conclusion}

The final lines of Table 2 indicate that the minimum energy states of the four TSAs 
all would be much flatter than observed (subject to the constraints that they remain HSLs).  
This leads me to reject the hypothesis that TSAs take the shape of lowest total energy $E$, 
subject to the constraint the they keep the same mass, density, and angular momentum, 
while remaining HSLs.  Therefore other processes must control the shapes of TSAs, 
involving angle of repose, for example.

\newpage

\section*{Appendix A: Derivation of the MOI of an HSL}

The value of the greatest moment of inertia of an Homogeneous Symmetric Lens derived in the unpublished manuscript by Levinson (2010) is in conflict with that published as Formula (60) of Kondrat'ev and Antonov (1993). The latter result is clearly in error, as it is based upon Formula (28) of Kondrat'ev and Antonov, which reduces to the wrong limits both for $h = 2R_C$ (sphere) and $h = R_C$ (hemisphere). As Levinson's work is not generally available, we present our calculation of the greatest of the MOI of an HSL in this Appendix. 

The greatest moment of inertia of an HSL is double that of the corresponding segment of a sphere. The integration is most easily performed in cylindrical polar coordinates $(r,\theta,z)$ (here and in Appendix B we do not use dagger superscripts for dummy variables as some of these variables are squared or raised to higher powers): 

\begin{equation}
				C = 2\int \rho r^2 dV 
	= 2 \int_{R_C-h}^{R_C}dz \int_0^{({R_C}^2-z^2)^{1/2}} dr\int_{0}^{2\pi} \rho r^3 d\theta .
\end{equation} 


\noindent The general result  (for $h < R$) of this integration yields:  

\begin{equation}
C = \frac{2\pi}{3}  \rho h^3 \left( 2R_C^2 - \frac{3}{2} R_Ch + \frac{3}{10} h^2 \right). 
\end{equation}

\noindent Note that the last 2 terms in Formula (14) are 3/2 as large as in Formula (60) of Kondrat'ev and Antonov (1993). Substituting $R_C = (R_E^2 +h^2)/(2h)$ into Formula (14) gives the result noted in Formula (3). 

\section*{Appendix B: Core-Mantle Model}

The main text of this paper treats only homogeneous bodies.  However, 
some models of TSAs feature a loose regolith on top of a solid core.  
For completeness, this Appendix generalizes the HSL model to a 
Core-Mantle Model (CMM), including a homogeneous spherical core at the center.  

Let a sphere of uniform density $\delta$, radius $R_S \le h$, 
and mass $M_S = 4\pi\delta R_S^3/3$ be centered at the geometric center 
of a HSL of density $\rho$ (see upper panel of Figure 9).  
Then the total density of the core is $\rho +\delta$.  
To represent a sparser core, or even a central void, 
$\delta$ may be negative, as long as the central density $\rho +\delta$ is not.  

\begin{figure}
\includegraphics[width=5.25in]{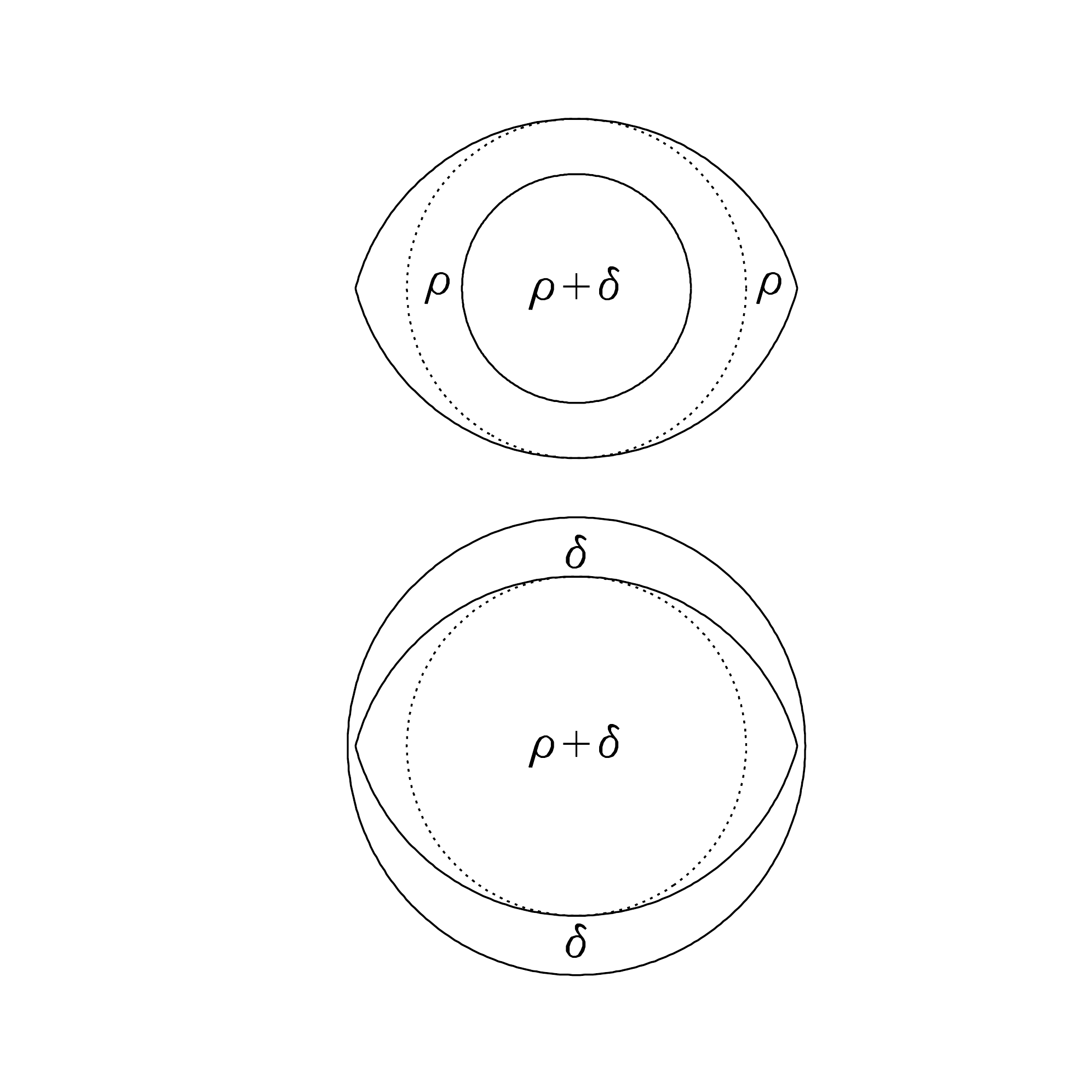}
\caption{ Upper panel: Polar cross-section of the Core-Mantle Model of a Top-Shaped Asteroid.  
Lower panel:  Polar cross-section of the Inverted Core-Mantle Model. 
The dotted circles represent limits of integration. }
\end{figure}

The center of mass and axis of greatest inertia of the HSL 
do not change, but its greatest moment of inertia now becomes 
\begin{equation}
		C_+ = \pi\rho h[10R_E^4 +5R_E^2 h^2 +h^4]/30 +8\pi\delta R_S^5/15 
\end{equation}
from Formula (3), while its equatorial moments of inertia are now 
\begin{equation}
                A_+ = B_+ = \pi\rho h[10R_E^4 +15R_E^2 h^2 +7h^4]/60 +8\pi\delta R_S^5/15 .
\end{equation}
from Formula (4).  Again, $A_+ \le B_+ \le C_+$, while $A_+ +B_+ \ge C_+$, as required.  
Similarly, the spin angular momentum of the CMM is $J_+ = C_+\omega$, 
and its rotational kinetic energy is $E_{R+} = C_+\omega^2/2 = J_+^2/(2C_+)$.  

The self-gravitational energy $E_G$ of the CMM 
can be expressed as the sum of three terms:  $E_G = E_\rho +E_\delta +E_{\rho\delta}$.  
Here $E_\rho$ is just the self-gravitational energy of the HSL, given by Formula (9). 
Likewise, $E_\delta$ is the self-gravitational energy of the central sphere 
of density $\delta$, given by $E_\delta = \frac{-3}{5}GM_S^2/R_S = 
\frac{-3}{5}G[\frac{4\pi}{3}\delta R_S^3]^2/R_S = -16\pi^2 G\delta^2 R_S^5/15$.  
Finally, $E_{\rho\delta}$ is the energy of the gravitational interaction between the HSL and the central 
sphere of density $\delta$, equal to minus the energy needed to separate them a great distance apart.  

The interaction energy $E_{\rho\delta}$ can be expressed in two ways:  either as 
\begin{equation}
				\int \delta U_\rho dx dy dz ,
\end{equation}
the integral of the gravitational energy of the central sphere of density $\delta$ 
in the potential field $U_\rho$ of the HSL, taken over the volume of the core; or as 
\begin{equation}
                                \int \rho U_\delta dx dy dz ,
\end{equation}
the integral of the gravitational energy of the HSL in the potential field $U_\delta$ 
of the central sphere of density $\delta$, taken over the entire volume of the HSL.  
Note that the leading factor of 1/2 in Formula (8) 
does not appear in either Formulae (17) or (18) above.  

Formula (17) is problematic, because the potential field $U_\rho$ of the HSL is unknown.  
Formula (18) is more approachable, but the potential field $U_\delta$ of the central sphere 
of density $\delta$ must be divided into an internal field and an external field.  
The attraction of the central sphere of density $\delta$ on an Exterior point 
at a distance $r$ from its center is $GM_S/r^2 = \frac{4\pi}{3}G\delta R_S^3/r^2$ 
toward its center, so the corresponding exterior gravitational potential $U_\delta$ 
is just $-GM_S/r = -\frac{4\pi}{3}G\delta R_S^3/r$.  

The attraction of the central sphere of density $\delta$ on an Interior point 
at a distance $r$ from its center is $4\pi G\delta r/3$ toward its center.  
Then matching the interior and exterior potentials at the core-mantle boundary 
gives a parabolic profile $U_\delta = -2\pi G\delta[R_S^2 -r^2/3]$  
of the interior gravitational potential.  

Now we can write the interaction energy $E_{\rho\delta}$ as the sum of two terms, 
an interior term $E_I$ and an exterior term $E_O$.  The interior term is then  
\[
				E_I = \int \rho U_\delta dx dy dz 
			= 4\pi \rho \int_0^{R_S} -2\pi G\delta[R_S^2 -r^2/3]r^2 dr 
			= -8\pi^2 G\rho\delta \int_0^{R_S} [R_S^2 r^2 -r^4/3] dr 
\] \begin{equation}
			= -8\pi^2 G\rho\delta [R_S^2 r^3/3 -r^5/15]_0^{R_S} 
			= -32\pi^2 G\rho\delta R_S^5/15 = -8\pi G\rho M_S R_S^2/5 , 
\end{equation}
where the integral is taken over the volume of the core.  

The exterior term $E_O$ is also sub-divided into two parts.  
The inner of these, $E_1$, is integrated from the surface of the core 
to the sphere $r = h$, indicated by the dotted circle in the upper panel of Fig. 9.  Then 
\[
				E_1 = \int \rho U_\delta dx dy dz
				= 4\pi \rho \int_{R_S}^h -GM_S rdr
\] \begin{equation}
				= -2\pi \rho GM_S [h^2 -R_S^2] 
			= -8\pi^2 G\rho\delta R_S^3 [h^2 -R_S^2]/3 .
\end{equation}
Note how $E_1$ vanishes if $R_S = h$.  

The outermost term $E_2$ of the interaction energy is the trickiest.  
First, in the ``northern'' hemisphere, define the latitude $\lambda$ 
as the angle between the body's equator plane 
and the ray from its center to the field point $\bf r$.  
Then the Law of Cosines gives 
\begin{equation}
		R_C^2 = r^2 +[R_C -h]^2 -2r[R_C -h]\cos(\lambda +90^\circ) 
			= r^2 +[R_C -h]^2 +2r[R_C -h]\sin\lambda .
\end{equation}

Solving Eq. (21) above for $r$ as a function of $\lambda$ 
leads to awkward limits of integration.  However, solving Eq. (21) 
for $\lambda$ as a function of $r$ is much more tractable:  
\begin{equation}
		\sin\lambda = \frac{2hR_C -h^2 -r^2}{2r[R_C -h]} . 
\end{equation}
As expected, Formula (22) above gives $\lambda = 90^\circ$ 
at the inner limit of integration $r = h$, 
while it gives $\lambda$ = 0 at the outer limit $r = R_E$.  

The $E_2$ term can be evaluated using integration in spherical coordinates $(r,\lambda,\phi)$.  
The integration over longitude $\phi$ is trivial, 
but the integral over $\lambda$ must be doubled to cover both ``hemispheres'':  
\begin{equation}
				E_2 = \int \rho U_\delta dx dy dz 
	= 2\rho \int_{-\pi}^\pi d\phi \int_0^{\lambda(r)} \cos\lambda d\lambda \int_h^{R_E} -GM_S rdr
\end{equation} \[
		= -4\pi\rho GM_S \int_0^{\lambda(r)} \cos\lambda d\lambda \int_h^{R_E} rdr 
			= -4\pi\rho GM_S \int_h^{R_E} \frac{2hR_C -h^2 -r^2}{2R_C -2h} dr
\]
\[
		= \frac{-2\pi\rho GM_S}{R_C -h} \{ [2hR_C -h^2][R_E -h] -[R_E^3 -h^3]/3 \} 
		= \frac{-8\pi^2 G\rho\delta R_S^3}{3R_C -3h} \{ [2hR_C -h^2][R_E -h] -[R_E^3 -h^3]/3 \} .
\]
Note how $E_2$ vanishes if $R_E = h$.  

Figure 10 contours $E_I$, $E_1$, and $E_2$ from Formulae (19), (20), and (22) above, 
as well as the total interaction energy $E_{\rho\delta} = E_I +E_1 +E_2$, 
all as functions of the dimensionless aspect ratio $\eta \equiv h/R_E$, 
as well as of the dimensionless ratio $R_S/h$, both running from 0 to 1.  
All four functions are normalized by $-32\pi^2 G\rho\delta R_V^5/15$, 
the interaction energy between two coincident spheres of uniform density $\rho$ and $\delta$.  

\begin{figure}
\includegraphics[width=5.25in]{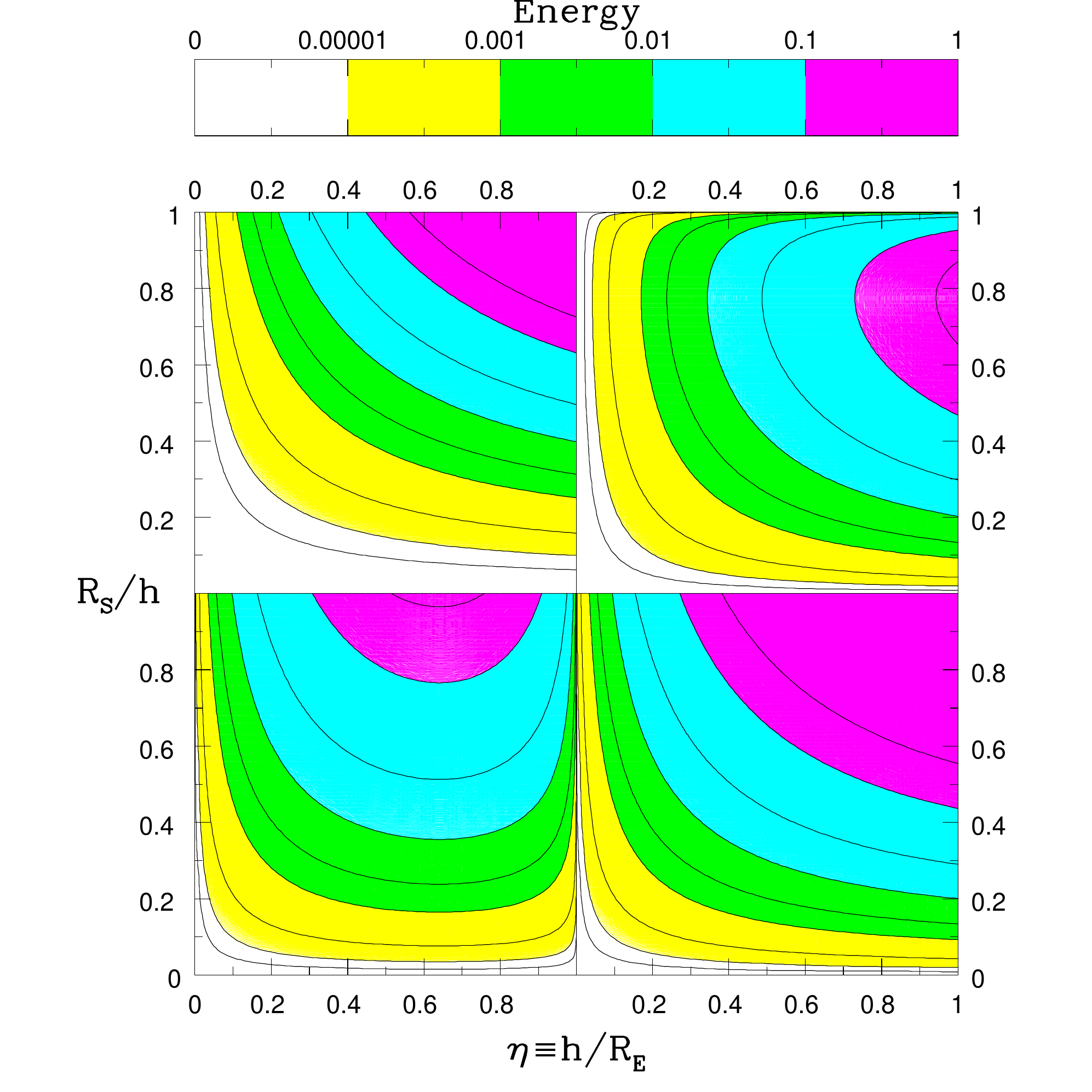}
\caption{ Contours of the interaction energy $E_{\rho\delta}$ 
between an HSL of uniform density $\rho$ and a smaller central sphere 
of uniform density $\delta$, as well as its component terms.  
Upper left panel:  $E_I$.  Upper right panel:  $E_1$.  
Lower left panel:  $E_2$.  Lower right panel:  $E_{\rho\delta} = E_I +E_1 +E_2$. 
All panels normalized by $-32\pi^2 G\rho\delta R_V^5/15$.  
Contour values:  0.000001, 0.00001, 0.0001, 0.001, 0.003, 0.01, 0.03, 0.1, and 0.2 .  }
\end{figure}


As expected, all four functions in Fig. 10 vanish as $R_S$ approaches zero, 
and the central sphere vanishes.  Furthermore, all four functions 
also vanish as $\eta$ approaches zero, and the CMM becomes a massless wafer.  
In addition, the term $E_1$ vanishes for $R_S = h$, 
while the term $E_2$ vanishes for $h = R_E$, both as expected.  

Note how $E_I$ and $E_{\rho\delta}$ look qualitatively similar, 
while $E_1$ and $E_2$ complement each other.  
$E_1$ reaches a maximum of $\sim$0.2323 at $\eta$ = 1, $R_S/h \approx$ 0.77, 
while $E_2$ reaches a maximum of $\sim$0.2225 at $\eta \approx$ 0.64, $R_S/h$ = 1.  
From $E_{\rho\delta}$, $E_\rho$, and $E_\delta$, one can find the total self-gravitational 
energy $E_G = E_\rho +E_\delta +E_{\rho\delta}$ of any Core-Mantle Model of a TSA. 
Finally, adding its rotational energy $E_{R+}$ gives its total energy $E = E_G +E_{R+}$.

\subsection*{Inverted Core-Mantle Model}

For completeness, consider also an Inverted Core-Mantle Model (ICMM) where a HSL of density $\rho$ 
is centered inside a larger sphere of uniform density $\delta$ (see lower panel of Fig. 9).  
As long as $\rho$ is positive, the center of mass, moments of inertia, 
and rotational kinetic energy do not change from Formulae (3), (4), and (6).  

However, if $\rho$ is negative, representing a denser mantle over a sparser core, 
the center of mass still does not change, but Formula (3) then gives the moment 
of {\it least} inertia, while Formula (4) gives the moments of greatest inertia.  
Then the ICMM rotates about an axis through its plane of symmetry, 
and its angular momentum becomes $J_+ = A_+\omega$, 
while its rotational kinetic energy becomes $E_{R+} = A_+\omega^2/2 = J_+^2/(2A_+)$.  

Again, the potential field of an HSL is unknown, 
but the interior potential of the sphere is $U_\delta = -2\pi G\delta[R_S^2 -r^2/3]$.  
Then the interaction energy $E_{\rho\delta}$ is just given by Formula (18) again.  
As with $E_O$, now $E_{\rho\delta}$ can be split into two parts:  

\[
E_{\rho\delta} = -4\pi G\rho\delta \int_{-\pi}^\pi d\phi \int_0^{\pi/2} \cos\lambda d\lambda \int_0^h [R_S^2 -r^2/3] r^2dr
	-8\pi G\rho\delta \int_{-\pi}^\pi d\phi \int_0^{\lambda(r)} \cos\lambda d\lambda \int_h^{R_E} [R_S^2 -r^2/3] r^2dr
\] \[
= -8\pi^2 G\rho\delta \int_0^h [R_S^2 -r^2/3] r^2dr -16\pi^2 G\rho\delta \int_h^{R_E} \frac{2hR_C -h^2 -r^2}{2R_C -2h}[R_S^2 -r^2/3] rdr
\]
\[
   = -8\pi^2 G\rho\delta \int_0^h [R_S^2 -r^2/3] r^2dr -\frac{8\pi^2 G\rho\delta}{R_C -h}\int_h^{R_E} [2hR_C -h^2 -r^2][R_S^2 -r^2/3] rdr
\] \[
				= -8\pi^2 G\rho\delta \left[ R_S^2 r^3/3 -r^5/15 \right]_0^h
	-\frac{8\pi^2 G\rho\delta}{R_C -h} \left[ [2hR_C -h^2]R_S^2 r^2/2 +[h^2/3 -2hR_C/3 -R_S^2]r^4/4 +r^6/18 \right]_h^{R_E}
\] \[
				= -8\pi^2 G\rho\delta [ R_S^2 h^3/3 -h^5/15] 
\] \begin{equation}
-\frac{4\pi^2 G\rho\delta}{R_C -h} \left[ [2hR_C -h^2]R_S^2[R_E^2 -h^2] +[h^2/3 -2hR_C/3 -R_S^2][R_E^4 -h^4]/2 +[R_E^6 -h^6]/9 \right] . 
\end{equation}


Figure 11 contours $E_{\rho\delta}$ from Formula (24) above, 
again normalized by $-32\pi^2 G\rho\delta R_V^5/15$.  
The format is similar to Fig. 10, with some important differences; primarily, 
the vertical scale in Fig. 11 refers to the dimensionless ratio $R_E/R_S \le 1$ 
of the equatorial radius of the HSL to the radius of the larger sphere, 
while that in Fig. 10 refers to the dimensionless ratio $R_S/h \le 1$ 
of the radius of the spherical core to the polar radius of the larger HSL.  

Generally, Fig. 11 for $E_{\rho\delta}$ of the ICMM resembles 
the lower right panel of Fig. 10, for $E_{\rho\delta}$ of the CMM.  
In both cases, $E_{\rho\delta}$ vanishes along the bottom and left-hand axes, 
and peaks at $-32\pi^2 G\rho\delta R_V^5/15$ in the upper right-hand corner.  

\begin{figure}
\includegraphics[width=0.9\linewidth]{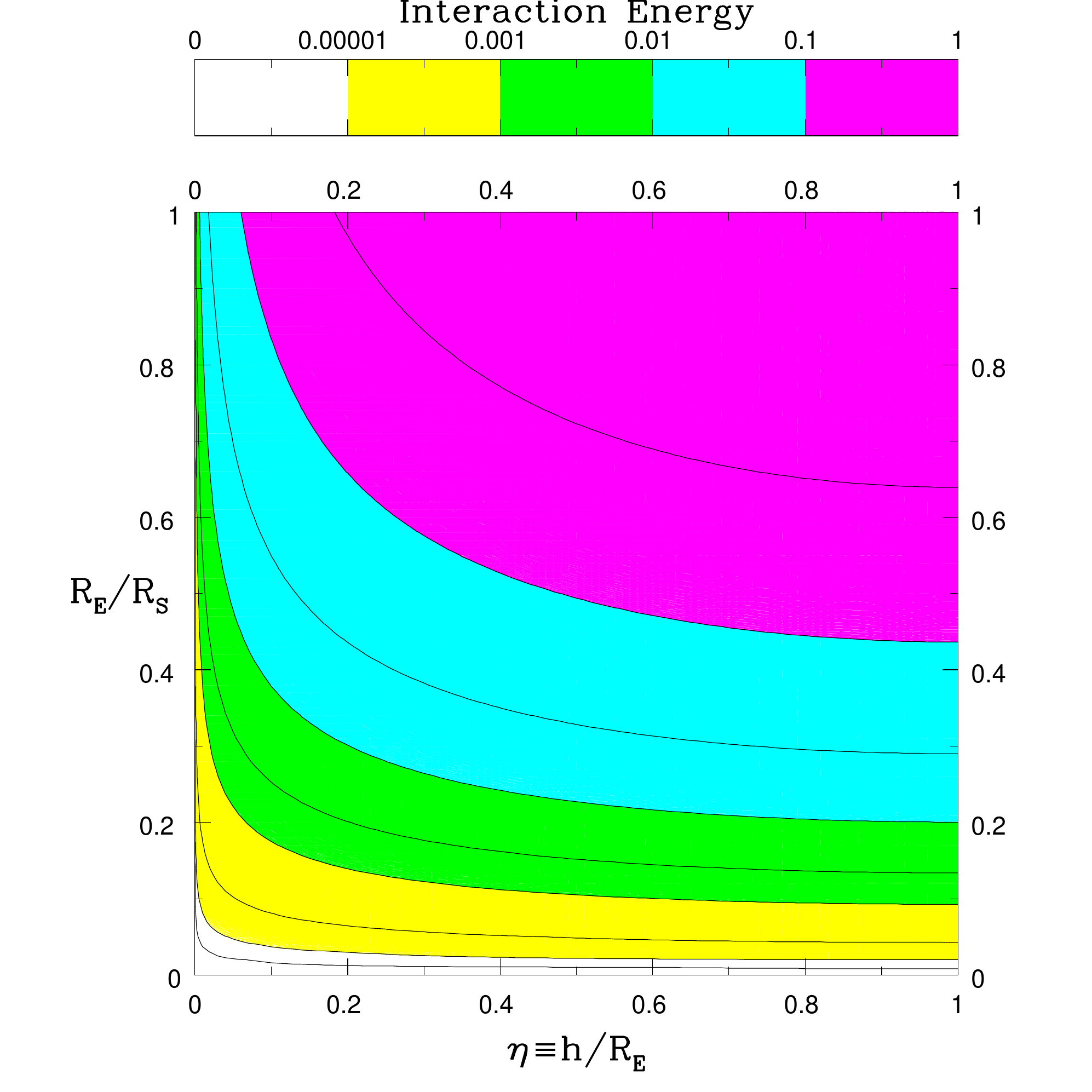}
\caption{ Contours of the interaction energy $E_{\rho\delta}$
(normalized by $-32\pi^2 G\rho\delta R_V^5/15$) 
between an HSL of uniform density $\rho$ 
and a larger central sphere of uniform density $\delta$. 
Contour values:  0.000001, 0.00001, 0.0001, 0.001, 0.003, 0.01, 0.03, 0.1, and 0.3 .  }
\end{figure}

From $E_{\rho\delta}$, $E_\rho$, and $E_\delta$, one can find the total self-gravitational
energy $E_G = E_\rho +E_\delta +E_{\rho\delta}$ of any Inverted Core-Mantle Model.  
Finally, adding its rotational energy $E_{R+}$ gives its total energy $E = E_G +E_{R+}$.

\newpage

\subsection*{Central potential of an HSL}

Finally, a similar integration gives the gravitational potential $U_0$ at the center of an HSL.  
Adapting Formula (7) to the origin gives just $U_0 = -G\rho\int dxdydz/r$.  
Again splitting the integral into two parts then gives
\[
			U_0 = -2G\rho\int_{-\pi}^\pi d\phi \int_0^{\pi/2} \cos\lambda d\lambda \int_0^h rdr 
			-2G\rho\int_{-\pi}^\pi d\phi \int_0^{\lambda(r)} \cos\lambda d\lambda \int_h^{R_E} rdr 
\] \[
			= -4\pi G\rho\int_0^h rdr -4\pi G\rho\int_h^{R_E} \left[ \frac{2hR_C -h^2 -r^2}{2R_C -2h} \right] dr 
\] \begin{equation}
			= -2\pi G\rho h^2 -\frac{2\pi G\rho}{R_C -h} \{ [2hR_C -h^2][R_E -h] -[R_E^3 -h^3]/3 \} . 
\end{equation}

Figure 12 graphs $U_0$ as a function of the aspect ratio $\eta \equiv h/R_E$, normalized by 
$-2\pi G\rho R_V^2$, the central potential of a sphere of radius $R_V$ and uniform density $\rho$.   
Note the almost hyperbolic form of the curve in Fig. 12, 
running from $U_0$ = 0 at $\eta$ = 0 up to $U_0$ = 1 at $\eta$ = 1, as expected.  

\begin{figure}
\includegraphics[width=5.25in]{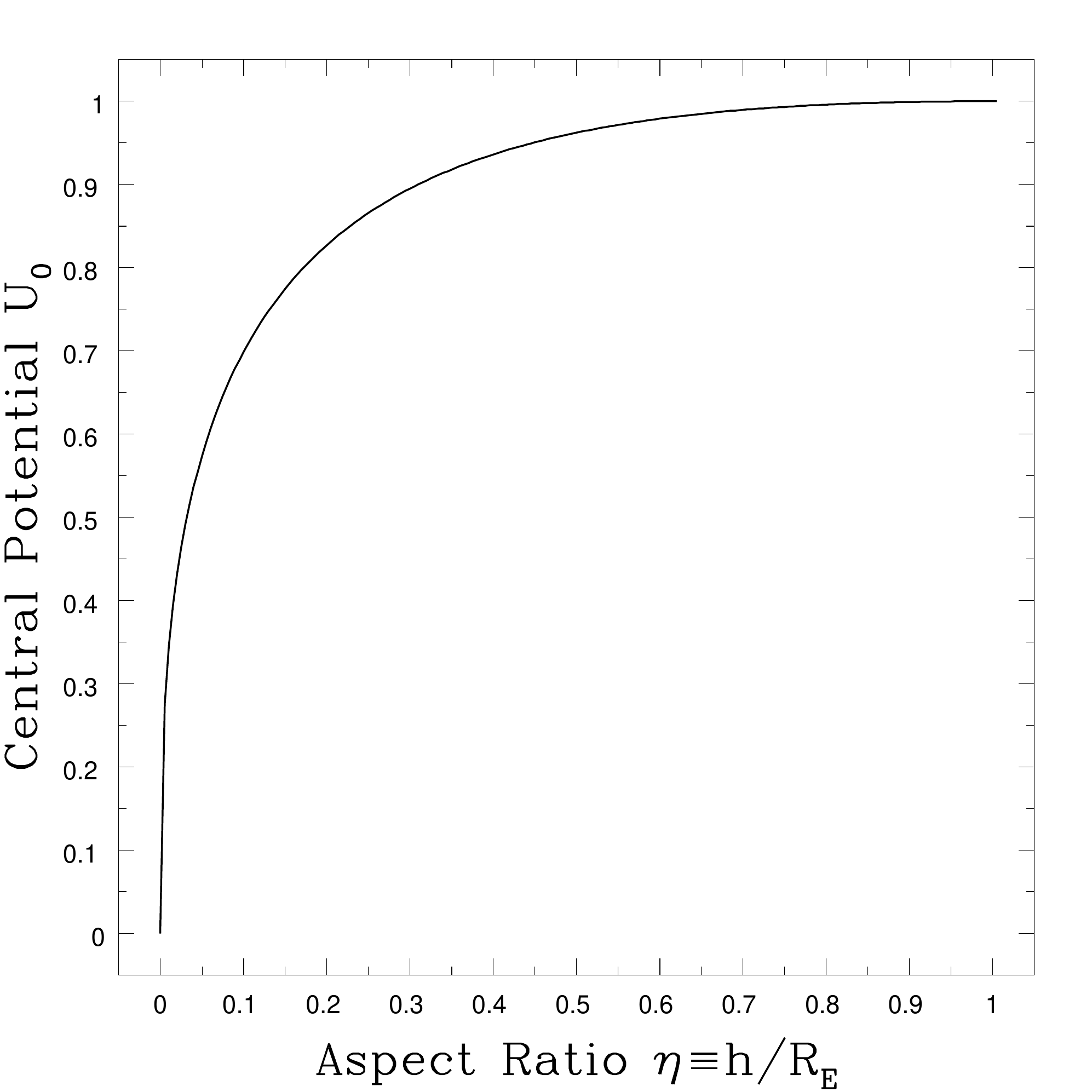}
\caption{ Central potential $U_0$ of an HSL, 
normalized by $-2\pi G\rho R_V^2$,  
as a function of its aspect ratio $\eta \equiv h/R_E$.  }
\end{figure}

\newpage

\begin{center}
                                ACKNOWLEDGEMENTS
\end{center}
{\bf The lead author of this paper (ARD) completed writing the manuscript on 2022 July 5 and sent it to two of his colleagues (Don Korycansky and JJL), shortly before suffering a stroke.  He was hospitalized the same day, and passed away July 21$^{\rm{st}}$ of complications of stage IV pancreatic cancer.  His colleagues Don Korycansky, Orkan Umurhan and the two listed co-authors (JJL and JLA) reviewed the manuscript, made some very minor corrections, and submitted it to Icarus as a single-author (ARD) work. One of the issues brought up the reviewer required considerable work by JJL and JLA to address, even though in the end they led to only minor changes in the manuscript, so JJL and JLA were added to the author list at that time.  We are grateful to Dan Scheeres for his 
thorough and thoughtful review. }   

\setlength{\parindent}{-0.3in}

\begin{center}
                                REFERENCES
\end{center}


Barnett, C. T., 1976. Theoretical modeling of the magnetic and gravitational fields \\
of an arbitrarily shaped three-dimensional body. {\it Geophysics} {\bf 41}, 1353--1364.




Broucke, R. A., 1995. Closed form expressions for some gravitational potentials: \\
Triangle, rectangle, pyramid and polyhedron. Pp. 1293--1309 in {\it Spaceflight Mechanics 1995:
\\ Proceedings of the AAS/AIAA Spaceflight Mechanics Conference}, R. J. Proulx, J. J. F. Liu,
\\ P. K. Seidelmann, and S. Alfano, editors. Univelt, Inc. Mostly republished in 1999
\\ as Pp. 321--340 in {\it The Dynamics of Small Bodies in the Solar System},
\\ B. A. Steves and A. E. Roy, editors. Kluwer, Inc.


Busch, M. W., S. J. Ostro, and 11 co-authors, 2011.  
Radar observations and the shape of near-Earth asteroid 2008 EV5.  
{\it Icarus} {\bf 212}, 649--660.  





Daly, M. G., and 36 co-authors, 2020.  Hemispherical differences 
in the shape and topography of asteroid (101955) Bennu.  
{\it Science Advances} {\bf 6} (41), eabd3649.  

Danby, J. M. A., 1962. {\it Fundamentals of Celestial Mechanics}. MacMillan.






Dobrovolskis, A. R., 1996.  Inertia of any polyhedron.  {\it Icarus} {\bf 124}, 698--704.  

Dobrovolskis, A. R., 2021.  Surface potential of a rotating duplex 
consisting of two conjoined spheres.  {\it Icarus} {\bf 358}, 114061.

Ivory, J., 1809. On the attractions of homogeneous ellipsoids.
{\it Phil. Trans. Roy. Soc.} {\bf 99}, 345--372.



Kellogg, O. D., 1929. {\it Foundations of Potential Theory}.
Springer, reprinted by Dover, 1953.

Kondrat'ev, B. P., 1989.  Dinamika Ellipsoidal'nykh Gravitiruiushchikh Figure \\ 
(Dynamics of Ellipsoidal Gravitating Figures).  272 pp. (in Russian).  Nauka, Moscow (in NASA ADS)

Kondrat'ev, B. P., 1993.  New methods in the theory of the Newtonian potential.  \\ 
Representation of the potential energy of homogeneous gravitating bodies by convergent series.  \\ 
{\it Astron. Zh.} {\bf 70}, 583--593 (in Russian).  Translated to English in {\it Astron. Rep.} {\bf 37}, 295--300.  

Kondrat'ev, B. P., and V. A. Antonov, 1993.  New methods in the theory of the Newtonian potential.  
\\ Potential energy of homogenous lens-shaped bodies and segments of spheres.  \\ 
{\it Astron. Zh.} {\bf 70}, 594--609 (in Russian).  Translated to English in {\it Astron. Rep.} {\bf 37}, 300--307.  

Kondrat'ev, B. P., 2003.  The Theory of Potential and Equilibrium Figures.  
\\ 624 pp. (in Russian).  Institute for Computer Research, Moscow-Izhevsk.  

Kondrat'ev, B. P., 2007.  Potential Theory:  New Methods and Problems with Solutions.  515 pp. 
(in Russian). \\ MIR, Moscow.  



Lauretta, D. S., and 29 co-authors, 2019.  
The unexpected surface of asteroid (101955) Bennu.  
{\it Nature} {\bf 568}, 55--60.  


Levinson, D., 2010.  Inertia properties of a segment of a uniform solid sphere.  Unpublished ms.  



MacMillan, W. D., 1930. {\it The Theory of the Potential}.
McGraw-Hill, reprinted by Dover, 1958.







Newton, I., 1687. {\it Philosophi{\ae} Naturalis Principia Mathematica}. Streater, London.

Okabe, M., 1979. Analytical expressions for gravity anomalies due to homogeneous polyhedral bodies 
\\ and translations into magnetic anomalies.  {\it Geophysics} {\bf 44}, 730--741.  


Ostro, S. J., and 15 co-authors, 2006.  
Radar imaging of binary near-Earth asteroid (66391) 1999 KW4.  
{\it Science} {\bf 314}, 1276--1280.  

Paul, M. K., 1974. The gravity effect of a homogeneous polyhedron for 
three-dimensional interpretation. \\ {\it Pure Appl. Geophys.} {\bf 112}, 553--561.  

Poh\'{a}nka, B., 1988. Optimum expression for computation of the gravity field of 
a homogeneous polyhedral body. \\ {\it Geophys. Prospecting} {\bf 36}, 733--751.  




Ramsey, A. S., 1940. {\it An Introduction to the Theory of Newtonian Attraction}.
Cambridge U.


Seidov, Z. F., 2000a.  Gravitational energy of simple bodies:  
the method of negative density.  arXiv:astro-ph/0003239.  

Seidov, Z. F., 2000b.  Gravitational potential energy of simple bodies:  
the homogeneous bispherical concavo-convex lens.  arXiv:astro-ph/0003233.  










Waldvogel, J., 1979. The Newtonian potential of homogeneous polyhedra. \\
{\it Z. Angew. Mathe. Phys.} {\bf 30}, 388--398.


Watanabe, S., and 87 co-authors, 2019.  
Hayabusa2 arrives at the carbonaceous asteroid 162173 Ryugu 
— A spinning top–shaped rubble pile.  
{\it Science} {\bf 364}, 268--272.  


Werner, R. A., 1994. The gravitational potential of a homogeneous polyhedron 
or don't cut corners. \\ {\it Cel. Mech. Dyn. Astron.} {\bf 59}, 253--278.  

Werner, R. A., and D. J. Scheeres, 1995. Exterior gravitation of a polyhedron derived
and compared with harmonic and mascon gravitation representations of asteroid 4769 Castalia.
{\it Cel. Mech. Dyn. Astron.} {\bf 65}, 313--344.




\end{document}